\documentclass[twocolumn,showpacs]{revtex4}
\usepackage{graphicx}
\usepackage{dcolumn}
\usepackage{amsmath}
\usepackage{bm}
\begin{document}

\preprint{HEP/123-qed}

\title[Fluctuations in Granular Media]{Turbulent-like Fluctuations 
in Quasistatic Flow of Granular Media}

\author{Farhang Radjai}
\affiliation{LMGC, CNRS-Universit\'e Montpellier II, 
Place Eugène Bataillon, 
34095 Montpellier cedex, France.}

\author{St\'ephane Roux}
\affiliation{Laboratoire ``Surface du Verre et Interfaces'', 
CNRS-Saint Gobain, 39 Quai Lucien Lefranc,
93303 Aubervilliers Cedex, France.}

\date{\today}

\begin{abstract}
We analyze particle velocity fluctuations in a simulated granular 
system subjected to homogeneous quasistatic shearing.  
We show that these fluctuations share the following scaling  
characteristics of fluid turbulence in spite of 
their different physical origins:  
1) Scale-dependent probability distribution  
with non-Guassian broadening at small time scales; 
2) Power-law spectrum, reflecting long-range 
correlations and the self-affine nature of the fluctuations;
3) Superdiffusion with 
respect to the mean background flow. 
\end{abstract}


\maketitle


The key role of fluctuations in quasistatic (QS) flow 
of granular media has been noted by several authors 
refering basically to stress fluctuations in time or 
the inhomogeneous distribution of forces in 
space\cite{miller,radjai}.     
Amazingly, few studies have been reported about 
the fluctuations of particle velocities under homogeneous strain 
conditions. These fluctuations have been observed to  
occur in a correlated fashion, though their 
scaling properties have not yet been analyzed\cite{misra,kuhn}.     
Other recent studies concern mainly  Couette 
flows where the strain is localized in the vicinity 
of the inner rotating cylinder\cite{losert_bocquet}. 


Analogies may be drawn with driven multiphase media  
dominated by steric interactions at the microscopic scale.    
Rheologically oriented studies of foams, for example, 
reveal large-scale cell rearrangements and vortex-like structures 
of the velocity field that control the complex flow 
behavior of foams\cite{durian,debregeas}. However, 
the (capillary) elasticity of cell walls is an important 
factor that discourages a closer comparison with 
systems composed of stiff elements such as granular materials. 
Concentrated suspensions provide a closer analogy in this respect.  
There exists now convincing evidence based on particle tracking 
that non-Brownian particles in a sheared 
fluid at small Reynolds numbers undergo a (kinematic) 
diffusive motion beyond a 
concentration-dependent set-in time\cite{rouyer,marchioro}. 


A much more remote comparison can be made with 
the field of turbulence. 
There obviously the physics is fondamentally different 
from that governing granular media. 
Nevertheless, the rich body of work devoted to the
statistical analysis of the fluctuating part of the velocity field in
turbulence provides a suitable framework  
that can be applied in order to
characterize the analogous fluctuating part of the velocity field in a
granular medium. To be more concrete, let us consider the 
following key aspects of turbulent fluctuations\cite{frisch}: 
1) Non-Gaussian broadening 
of the probaility density functions (pdf's)  
of velocity differences as a manifestation 
of small-scale intermittency,
2) Multiscale organization 
of the velocity field reflected in its power-law spectrum, 
and 3) Anomalous diffusion pertaining to the 
Richardson regime. 
Although deeply rooted in fluid dynamics (Navier-Stokes equations, inertia 
regime), these scaling features {\em may}, in principal, prove to be  
relevant as well within a different physical context such as granular 
flows.  


Following this route, the objective of this paper is to show that 
particle velocity fluctuations in a QS granular flow, simulated by means 
of the molecular dynamics method, exhibit indeed strikingly similar features. 
A strict analogy with turbulence makes certainly not much sense 
because of a drastically different physics that underlies 
these fluctuations. 
In particular, inertia effects are basically irrelevant in 
a QS granular flow, and frictional and hard-core 
inelastic interactions between particles have little in common with molecular 
interactions in a fluid.   
But, precisely because of these dissimilarities, the observed analogy in 
terms of scaling properties 
is quite nontrivial and it might lead to new insights in both fields.  
In the following, we first describe the simulated granular 
system and our procedures of data analysis. 
Then, we present our main results focussing on the pdf's, correlations 
and diffusion, respectively.

\begin{figure}[t]
\includegraphics[width=6cm]{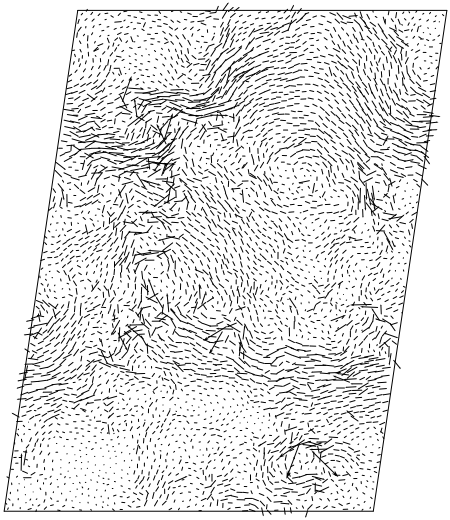}
\caption{\label{fig1} A snapshot of particle 
displacements $\delta {\bm s}^i $ with respect to the mean 
background flow.}
\end{figure}


The investigated granular model is a two-dimensional  
assembly of 4000 frictional discs with diameters uniformly  
distributed between $D_{min}$ and $D_{max}$ with $D_{max} = 3 D_{min}$. 
The particles interact through a stiff 
linear repulsive force as a function of mutual overlaps  
and the Coulomb friction law. The equations of motion for 
particle displacements and rotations are integrated by means of   
a predictor-corrector scheme\cite{allen}. 


An accurate evaluation of the statistics 
of fluctuations requires long-time homogeneous and steady shearing.       
However, ordinary wall-type boundary conditions  
induce a pronounced layering effect and the 
corners enhance the local frustrations 
whereby large strain and stress inhomogeneities arise 
when the box shape changes. System-size 
inhomogeneities may also occur due to  shear localization. 
In order to circumvent such unwanted effects, we used 
bi-periodic boundary conditions following a method 
similar in spirit to that devised by Parrinello 
and Rahman\cite{parrinello}. 


In our simulations, the gravity was set to zero and 
a confining pressure was applied along the $y$ direction.      
The width $L$ of the simulation cell was kept constant. 
The displacement field is decomposed 
into an affine displacement field ${\delta {\bm r}}^i 
\equiv (\delta r_x^i, \delta r_y^i)$ 
and a fluctuating field ${\delta {\bm s}}^i 
\equiv (\delta s_x^i,\delta s_y^i)$ of zero 
mean ($\langle{\delta {\bm s}}\rangle = 0$) carried by the particles $i$. 
The system is driven by imposing 
${\delta r_x}^i =  \delta t \gamma r_y^i$, where $\gamma$ is 
a constant shear rate and $\delta t$ is the time step. 
In other words, the $k=0$ Fourier mode of the
total strain is imposed, corresponding to 
a large scale forcing.  
This driving mode was applied on a dense packing  
prepared by isotropic compaction. 
The height $H$ of the packing increases (dilation) in 
the initial stages of shearing before 
a homogeneous steady state is reached  
where ${\delta H}/H  \equiv {\delta r_y}^i /r_y^i$ fluctuates around zero. 
The focus of this paper is the field 
${\delta {\bm s}^i}$ which corresponds to a  
spatially periodic motion of the particles with 
respect to the background flow $\delta{\bm r}^i$. 


Although our dynamic simulations involve the 
physical time, the inertial effects are negligibly small and the granular 
texture evolves quasistatically at time scales well below 
$\gamma^{-1}$. We normalize all times 
by $\gamma^{-1}$ so that the dimensionless time $t$ in what follows 
will actually represent the cumulative shear strain.  
We will also use the mean particle diameter $D$  
to scale displacements. As a result, the velocities will be scaled by 
$\gamma D$ and the power spectra in space by $(D^2 \gamma)^2$.
In our simulations 
the time step was $\delta t  \simeq 10^{-7}$ (in dimensionless units), 
and more than $2.10^7$ steps were simulated, corresponding to 
a total strain larger than  $2$. The solid fraction 
and the average coordination number in the steady state were 
about $0.8$ and $3.8$, respectively.


It is important to emphasize here that the 
granular nature of our system does not allow to apply 
exactly the same procedures of data analysis as in turbulence. 
Turbulence studies focus mainly  on 
velocity differences $\delta v$   measured at a fixed 
point of a fluid over a time interval $\tau$ or between two points 
separated by a distance $r$. In contrast, 
the particle-scale granular motion involves a 
{\em discrete} displacement field that is carried by 
individual {\em particles}. Thus, 
our natural framework is Lagrangian rather than Eulerian. These  
differences are certainly important for a strict one-to-one 
comparison, but here we
basically consider turbulence as a reference field from which we extract
tools to characterize granular fluctuations.

Another distinctive feature of granular flow is that, 
due to collisions, the velocities are 
discontinuous in time. As the positions are better behaved, we 
characterize the fluctuating motions of the particles  by ``tracer'' 
velocities defined from particle displacements $\delta {\bm s}^i$ by 
\begin{equation}
{\bm v}^i(t,t+\tau) =  \frac{1}{\tau} 
\int_t^{t+\tau} \delta {\bm s}^i (t') \ dt'     
\label{eqn1}
\end{equation}
where $\tau$ is the time resolution.  
Since we are concerned with steady flow, the statistical properties 
of $\bm v$ (pdf's, correlations) are independent of $t$  
(though, as shown below, they crucially depend on $\tau$).   
Hence, accurate statistics can be obtained by cumulating the 
data from different time slices of a single simulation running for a long 
time. 


Fig.~\ref{fig1} shows a snapshot of   
fluctuating velocities ${\bm v}^i$ 
for a short time lag $\tau = 10^{-7}$.
We see that large-scale well-organized displacements 
coexist with a strongly inhomogeneous distribution of amplitudes 
and directions on different scales. Eddy-like structures   
(though without the singular
vorticity concentrated in the core of these eddies) appear quite 
frequently, but they survive typically for strains $\tau$ less than 
$10^{-3}$.  After such short times, large-scale eddies break down 
and new (statistically uncorrelated) structures appear. This behaviour 
is radically different from turbulence where the eddies 
survive long enough to undergo a significant distortion due to fluid 
motion\cite{frisch}. 

\begin{figure}[t]
\includegraphics[width=4cm,angle=-90]{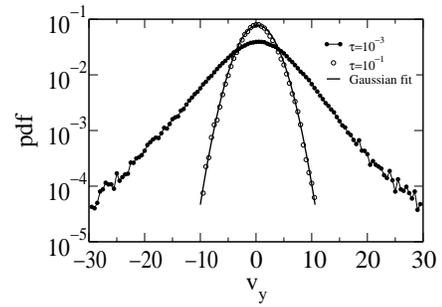}
\caption{\label{fig2} The pdf's of the y-components of fluctuating 
velocities for two different integration times: 
$10^{-3}$ (broad curve) and $10^{-1}$ (narrow curve). The latter 
is fitted by a Gaussian. The error bars are too small to be shown.}
\end{figure}

\begin{figure}[h]
\includegraphics[width=5cm]{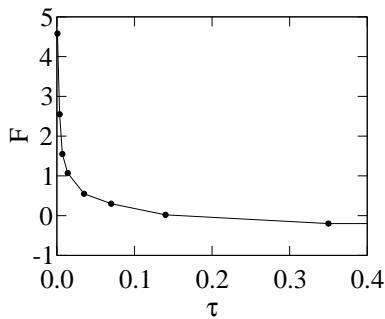}
\caption{\label{fig3} Flatness $F$ of the distribution  of velocity 
fluctuations as a function of the integration time $\tau$.}
\end{figure}


We consider now the pdf's of the 
components $v_x^i(t,t+\tau)$ and $v_y^i(t,t+\tau)$  
as a function of time resolution $\tau$. 
The pdf's of $v_y^i$ are shown in Fig.\ref{fig2} for a short integration 
time $\tau = 10^{-3}$, and for a long integration time $\tau = 10^{-1}$. 
The pdf has changed from a nearly Gaussian 
shape at large $\tau$ to a non-Gaussian shape with broad stretched 
exponential tails extending nearly to the center of the distribution       
at small $\tau$. We found no simple form allowing to fit 
the non-Gaussian pdf over the whole velocity range. 

In order to characterize this non-Gaussian broadening of the pdf's 
as a function of $\tau$, we calculated the flatness 
$F = \langle v_y^4 \rangle / \langle v_y^2 \rangle -3$, which is 
zero for a Gaussian distribution and $3$ for a purely 
exponential distribution. The values of $F$ as a function of $\tau$, shown 
in Fig.\ref{fig3}, are consistent with zero at large $\tau$ ($\tau > 0.2$) 
and rises to $5$ for our finest time resolution 
($\tau= 10^{-7}$). A strictly similar behaviour was observed for the 
component $v_x$. 

The broadening of the exponential tails of 
the pdf's at increasingly smaller 
scales is a hallmark of fully developped turbulence (for velocity 
differences)\cite{vincent}.  
It is attributed to  the phenomenon of intermittency, 
i.e. strong localized energy transfers at small scales. 
In a QS granular flow, 
the basic physical mechanism underlying the fluctuations is the mismatch 
of the uniform strain field applied at the boundaries or in the bulk, 
with mutual exclusions of the particles. As a result, 
the local strains deviate from the mean (global) strain. 
The observation of a transition toward a Gaussian
distribution for large time lags is a sign of loss of correlation and/or
exhaustion of large fluctuations in the increment of displacement which
occur at different times. 
Unfortunately, the rich multifractal scaling 
of velocity fluctuations is out of reach within the present investigation 
due to demanding statistics\cite{benzi}. 
The analogy with turbulence, however,  
suggests further study along such routes.

\begin{figure}[t]
\includegraphics[width=4cm,angle=-90]{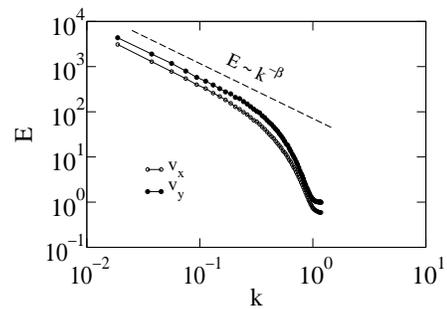}
\caption{\label{fig4} Averaged power spectrum of 
the $x$ and $y$ components of the fluctuating velocity field 
over one time step for one-dimensional cross sections along  
the mean flow.}
\end{figure}

In order to quantify the extent of these correlations, we estimated 
the power spectrum $E$ of velocity fluctuations  
both along and perpendicular to the flow and at different times. 
The Fourier transform was performed over the fluctuating 
velocity field defined on a 
fine grid by interpolating the velocities from particle centers. 
The power spectra were quite similar along 
and perpendicular to the flow, and for different 
snapshots of the flow. The averaged one-dimensional spectrum 
is shown in Fig.\ref{fig4}. It has 
a clear power-law shape $k^{-\beta}$ ranging from the 
smallest wavenumber $k=D/L$, corresponding to scale $L$, up  
to a cut-off around $k=0.5$, corresponding to nearly 
two particle diameters.
The exponent is $\beta \simeq 1.24 \simeq 5/4$ over one decade 
(to be compared with the exponent $5/3$ as a hallmark of 3D turbulence 
for the spectrum of velocity differences). This
means in practice that the fluctuating 
velocity field is self-affine with a Hurst exponent
$H = (\beta-1)/2= 0.12$\cite{feder}. 

Due to the peculiar behavior of the
velocity field (discontinuous in time), one might 
expect that the power spectrum is sensitive to the time resolution 
$\tau$. However, we checked that the value of $\beta$ is 
independent of $\tau$. It is also noteworthy that,  
the presence of long range  correlations in displacements, reflected 
in the value of $\beta$, is in strong contrast with the observed 
correlation lengths of nearly $10D$ for contact forces\cite{radjai}.

\begin{figure}[th]
\includegraphics[width=4cm]{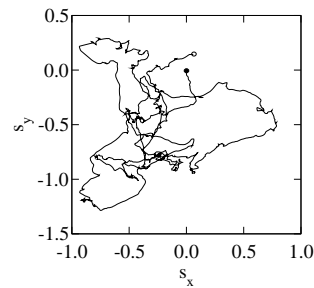}
\caption{\label{fig5} Diffusive trajectory of one 
particle with respect to the mean 
background flow for a cumulative shear strain of 2.}
\end{figure}

\begin{figure}[h]
\includegraphics[width=4cm,angle=-90]{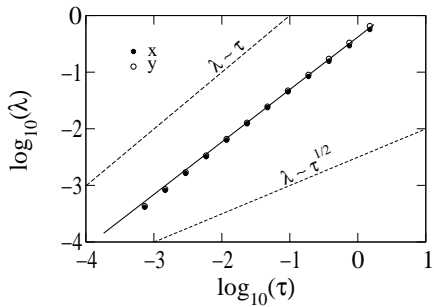}
\caption{\label{fig6} Root-mean-square relative displacements $\lambda$  
along $x$ and $y$ directions 
as a function  of time $\tau$ fitted by a   
power law of exponent $0.9$ (straight solid line). 
As a guide to the eyes, a power law of exponent $1/2$, corresponding to 
normal diffusion, and the line $\lambda \propto \tau$ are 
shown as well (dashed lines). The error bars are too small 
to be shown.}
\end{figure}

The long-time behaviour may be studied by considering 
the effective diffusion of the particles.  Normal diffusion 
implies that the root-mean-square (rms) relative displacements $\lambda$ 
in a given direction varies in proportion 
to the square root of time. 
In 3D fluid turbulence at high Reynolds numbers, the long-time 
pair diffusivity of suspended particles is anomalous, following 
the Richardson law $\lambda \propto \tau^{3/2}$.
This superdiffusion law reflects the Kolmogorov-Obukhov velocity spectrum 
$v_\lambda \propto \lambda^{1/3}$, where $v_\lambda = \lambda(\tau) /\tau$ 
is the characteristic velocity difference over 
a distance $\lambda$\cite{frisch,isichenko}. 

We analyzed the {\em kinematic} diffusion of single particles 
in our QS granular flow. 
One example of a single particle trajectory 
with respect to the background strain 
is shown in Fig.\ref{fig5}. We see that  the
fluctuating displacement is of the order of the mean
particle diameter for a strain of the order of unity. Fig.\ref{fig6} shows 
the rms relative displacement $\lambda(\tau)$ of all particle pairs 
initially in contact, as a function of time along $x$ and $y$ directions. 
This clearly corresponds to a superdiffusion behaviour, 
\begin{equation}
\lambda \propto \tau^\alpha,
\label{eqn2}
\end{equation} 
with $\alpha \simeq 0.9$ for both components over nearly $3$ decades 
of strain. Particle self-diffusivities exhibit a similar law. 
Since large-scale structures are short lived, 
anomalous diffusion scaling in our granular system 
can not be solely attributed to velocity correlations.  
This behaviour reveals, above all, the long-time 
configurational memory of a granular medium in QS flow. 

We note that Fig.\ref{fig6} shows no anisotropy for the diffusion. 
In fact, the steric exclusion effects dominate over the 
large-scale strain field for small diffusive displacements (one 
particle diameter, corresponding to $2\%$ of the cell size 
in our simulations), and hence the anisotropy may be weak 
at such scales.   
Moreover, we observe no crossover to normal scaling 
within the investigated strain range.  We cannot
exclude that for larger strains, when two particles are more widely
separated, a normal diffusion law is recovered.  However, this would
require extremely long computations, and, experimentally, strains of
the order of unity are already above standard tests.   


In summary, we 
analyzed fluctuating particle displacements with respect to the 
background quasistatic shear flow in a model 
granular medium. These fluctuations were shown to 
have the following scaling charactersitics: 
1) The pdf's undergo a transition from streched exponential to
gaussian as the time lag is increased; 
2) The power spectrum (in space) of the velocity field obeys a
power law, reflecting long range 
correlations and the self-affine nature of the fluctuations;
3) The fluctuating displacements have a superdiffusive character. 
These observations contradict somehow the conventional wisdom 
which disregards kinematic fluctuations in macroscopic modelling 
of plastic flow in granular media. Several basic aspects of 
quasistatic granular flow (elementary representative volumes, 
mean field approximation, memory effects, mixing) are thus concerned by 
these findings\cite{roux}.  

There appears an evident analogy with the scaling features of 
turbulence that was also discussed throughout this paper.
Of course, this analogy 
does not imply that a QS granular flow can be considered as turbulent in the 
standard sense of fluid dynamics. 
In particular, because of the fundamentally 
different origin of granular fluctuations, a direct reference to 
the underlying physics of turbulence can be 
misleading. 
But, the observed analogy in terms of scaling characteristics 
is consistent enough 
to upgrade kinematic fluctuations in quasistatic 
granular flows to the rank of a 
systematic phenomenology which could be coined by the term 
``granulence'' as compared to ``turbulence'' in fluids. 

Interestingly, this analogy works with three-dimensional 
turbulence although our data concern a two-dimensional granular flow.  
The energy cascade in turbulence is governed by inertia, 
and two- and three-dimensional systems do differ
significantly in this respect.  In quasistatic granular flow, 
the fluctuating velocity field is a consequence of the 
geometrical compatibility of the strain with  
particle arrangements, and dissipation is mainly governed  
by friction at the particle scale.  The difference
between two- and three-dimensional systems may thus be less 
crucial, but it was not investigated in the present work.  
Quite independently of its    
physical origins, this analogy is suggestive 
enough to be used as a promising strategy 
towards a refined probabilistic description of granular 
flow in the plastic regime.

It is a great pleasure to acknowledge  N. Schorghofer for 
his careful reading of the manuscript and valuable comments 
about this work, as well as J.E. Wesfreid and T. Witten for 
very interesting discussions.  


\newpage

\newpage

\newpage

\newpage

\newpage


%
\end{document}